\documentclass[aps,prd,preprint,groupedaddress]{revtex4-1}
\usepackage[dvips]{graphicx}
\usepackage{color}
\usepackage{array}
\usepackage{caption}

\def\eq{equation}
\def\eqn{eqnarray}

\begin{document}

\title{Highly-collimated, magnetically-dominated jets
\\ around rotating black holes}
\author{Zhen Pan}
\email{zhpan@ucdavis.edu}
\affiliation{Department of physics, University of California, Davis}
\author{Cong Yu}
\email{cyu@ynao.ac.cn} \affiliation{Yunnan Observatories, Chinese
Academy of Sciences, Kunming, 650011, China \\
Key Laboratory for the Structure and Evolution of Celestial
Object, Chinese Academy of Sciences, Kunming, 650011, China}
\date{\today}

\begin{abstract}
In this paper, we propose a general method for perturtative solutions
to Blandford-Znajek mechanism. Instead of solving the nonlinear
Grad-Shafranov equation directly, we introduce an alternative way to
determine relevant physical quantities based on the horizon
boundary condition and the convergence requirement.
Both the
angular velocity $\Omega$ of magnetic field lines, the toroidal magnetic field $B^\phi$ and the total electric current $I$ are
self-consistently specified according to our method.
As an example, stationary axisymmetric and force-free jet models around rotating black holes are
self-consistently constructed according to the method we proposed. 
This jet solution distinguishes itself from prior
known analytic solutions in that it is highly collimated and
asymptotically approaches a magnetic cylinder. This jet solution
is helically twisted, since toroidal magnetic field is generated
when the black hole spin is taken into account. 
For a given magnetic flux
threading the black hole, the jet power and energy extraction rate
of the collimated jet are compared with previous solutions.
We find that our new solution agrees better with current
state-of-the-art numerical simulation results. Some interesting
properties of the collimated jet and effects of field line
rotation on the jet stability are also briefly discussed.
\end{abstract}


\maketitle

\section{Introduction}
Many high-energy astrophysical objects, such as active galactic
nuclei (AGNs) and gamma-ray bursts (GRBs) as well as
ultra-strongly magnetized neutron stars (magnetars) involve
relativistic magnetically-dominated plasma. Under such
circumstances, magnetic fields play crucial roles in the dynamics
of these astrophysical scenarios, which can drive powerful
winds/jets from these astrophysical objects. It is widely accepted
that, in these objects, the magnetic energy density conspicuously
exceeds the thermal and rest mass energy density of particles. The
force-free electrodynamics behave well in such extreme
magnetically dominated scenarios as the less important terms, such
as the inertia and pressure, are entirely ignored. In the force
free electrodynamics the Lorentz force $F_{\mu\nu}J^{\nu}$
disappears. Based on the force-free electrodynamics, Blandford \&
Znajek (1977) studied an axisymmetric steady-state plasma
surrounding a spinning black hole and proposed that the rotation
energy of a Kerr black hole could be extracted via the action of
force free electromagnetic fields, in the form of Poynting flux
via magnetic field lines penetrating the central black hole. This
BZ mechanism is proved to be one of the most powerful energy
releasing processes in our universe and it is one of promising
candidates as the central engine of AGNs and GRBs.

The configuration of the ordered magnetic field around the black
hole has been discussed both in analytical and numerical studies.
A self-consistent description of the highly magnetized plasma
around strongly curved space-time of rotating black holes involves
the nonlinear Grad-Shafranov (GS) equation. First examples of the
force-free field configurations were constructed in
(\cite{Blandford1977d}, henceforth BZ77), in which perturbation techniques were
applied to get self-consistent field configurations. Some
following efforts, in which the angular velocity of magnetic field
is prescribed rather than self-consistently determined from the GS
equation, were made to model central regions of stationary
axisymmetric magnetosphere of black holes
(\cite{MacDonald1984y, Beskin1992b}). Due to the extreme
nonlinearity of the GS equation, there has been almost no any
further development in the analytic solution to the BZ mechanism.
General relativistic magnetohydrodynamics (GRMHD) and
magnetodynamics (GRMD) simulations, however, provide us an
opportunity to look into the nature of the BZ mechanism. GRMHD
simulations of black hole accretion system show that the
perturbative split monopole solution is consistent with the
numerical results in the low-density polar regions
(\cite{Mckinney2004f, Mckinney2005, Mckinney2006}). GRMD
simulations suggest that the split monopole solution is accurate
and stable for slow rotating black holes ($a\ll1$), where $a$ is
the specific momentum of the Kerr black hole, and is also a rather
good approximation for even fast rotating black hole ($a=0.9$)
(\citep{Komissarov2001, Komissarov2002c, Komissarov2004, Komissarov2004e}).
The question about the magnetic field configuration in the
vicinity of a black hole still remains an open issue. Recent
numerical investigation (\cite{Penna2013a}) shows that the
standard split monopole jet power is about 60\% larger than the
GRMHD simulation results, which indicates that the split monopole
jet model may not account for the simulation results properly. In
this paper, we propose a highly collimated jet model, which can
reduce the jet power to be more consistent with recent
simulations.

There are some concerns about stabilities of jet launched by the
BZ mechanism
(\cite{Li2000, Lyutikov2006e,Giannios2006}), but many
numerical simulations imply that the jet is stable. Especially,
three-dimensional GRMHD simulations have been performed to
investigate the stability of relativistic jets and no instability
was discovered (\cite{Mckinney2009a, Mckinney2013}). The
possible origin for the discrepancy is that analytical work, which
applies the Kruskal-Shafranov (KS) criteria to the highly
magnetized magnetosphere, does not consider the stabilizing
effects present in simulations, including field rotation, gradual
shear, a surrounding sheath, sideways expansion and non-linear
saturation (\cite{Mckinney2009a}). Including the field rotation,
the split monopole solution was analytically proved to be stable
against screw unstable modes satisfying the KS criteria
(\cite{Tomimatsu2001}). In this paper, we also study stabilities
of the new collimated jet solution.

This paper is organized as follows: basic equations governing
stationary axisymmetric force-free fields around Kerr black holes
are introduced in section 2. In section 3 we describe a
perturbative approach to obtain self-consistent highly collimated
jet solutions. Physical properties, such as energy extraction
rate, stability of the jet solution, are discussed and compared
with previous known solutions in section 4. Discussions are given
in section 5.

\section{Stationary Axisymmetric Force-Free Fields around Kerr Black Holes}

We adopt the Kerr-Schild coordinate (horizon penetrating,
\cite{Mckinney2004f}), in which the line element is
\[
ds^2 = -\left( 1-\frac{2r}{\Sigma} \right)dt^2 + \left( \frac{4
r}{\Sigma} \right) dr dt + \left(1+\frac{2r}{\Sigma} \right) dr^2
+ \Sigma d\theta^2 - \frac{4 a r \sin^2\theta}{\Sigma} d\phi dt
\]

\begin{\eq}
- 2 a \left(1+\frac{2r}{\Sigma}\right) \sin^2\theta d\phi dr  +
\sin^2\theta \left[\Delta  + \frac{2r (r^2 + a^2) } {\Sigma}
\right] d\phi^2 \ ,
\end{\eq}
where $\Sigma=r^2+a^2\cos^2\theta$, $\Delta=r^2-2r+a^2$, and
$\sqrt{-g}=\Sigma\sin\theta$ .

Since the magnetosphere around the black hole is magnetically
dominated, we adopt the force-free approximation, which ensures
that the electromagnetic field dominates over matter $T^{\mu\nu} =
T^{\mu\nu}_{\rm matter} + T^{\mu\nu}_{\rm EM}\approx
T^{\mu\nu}_{\rm EM}$. The energy-momentum tensor of the
electromagnetic field is $T^{\mu\nu}_{\rm EM} = F^{\mu\tau}
F^{\nu}_{\phantom{nu}\tau} - \frac{1}{4} \delta^{\mu\nu}
F^{\alpha\beta} F_{\alpha\beta}$, where the Faraday tensor is
defined as $F_{\mu\nu} = \partial_\mu A_\nu - \partial_\nu A_\mu$.
It is easy to prove that the energy-momentum conservation of
electromagnetic field is equivalent to the force-free condition
(\citep{Landau1980Classical}),
\begin{\eq}\label{energymomentum}
T^{\mu\sigma}_{\phantom{\mu\sigma};\sigma} = F^{\mu\nu} J_{\nu} =
0 \ .
\end{\eq}
The force-free condition implies the vanishing of the electric
field in the local rest-frame of the current, and thus
$^*F^{\mu\nu} F_{\mu\nu}=0$, where $^*F^{\mu\nu} \equiv
\frac{1}{2} \epsilon^{\mu\nu\rho\sigma} F_{\rho\sigma}$ is the
dual of the Faraday tensor. It is straightforward to prove that
$A_{\phi,\theta} \ A_{t,r} = A_{t,\theta} A_{\phi,r} \ $,
which indicates that $A_t$ is a function of $A_{\phi}$. We can
define the angular velocity of the magnetic field
$\Omega(r,\theta)$ as follows,
\begin{\eq}
- \Omega \equiv \frac{d A_t}{d A_{\phi}} =
\frac{A_{t,\theta}}{A_{\phi,\theta}} = \frac{A_{t,r}}{A_{\phi,r}}
\ ,
\end{\eq}
which is an unspecified function and will be determined
self-consistently in Section III.  For simplicity, we consider an
stationary and axisymmetric model, which implies that $F_{t\phi} =
0$ and the non-vanishing components of the antisymmetric Faraday
tensor $F_{\mu\nu}$ are as follows:
\begin{\eq}\label{poloidal}
F_{r\phi} = -F_{\phi r}=A_{\phi,r} \ , F_{\theta\phi} = -
F_{\phi\theta} = A_{\phi,\theta} \ ,
\end{\eq}
\begin{\eq}\label{electricfield}
F_{tr} = -F_{rt} = \Omega A_{\phi,r} \ , F_{t\theta}= - F_{\theta
t} = \Omega A_{\phi,\theta} \ ,
\end{\eq}
\begin{\eq}\label{toroidal}
F_{r\theta} = -F_{\theta r} = \sqrt{-g}B^{\phi} \ .
\end{\eq}
The above five non-zero components of $F_{\mu\nu}$ can be
specified in terms of three free functions $\Omega(r,\theta)$,
$A_{\phi}(r,\theta)$, $B^{\phi}(r,\theta)$.
According to the definition of the energy-momentum tensor, we can
further have that
$T^{\theta}_{\phantom{\theta}t} = -\Omega
T^{\theta}_{\phantom{\theta}\phi}$ and   $T^{r}_{\phantom{r}t} =
-\Omega T^{r}_{\phantom{r}\phi}$.    
With these two relations, the energy conservation and angular
momentum conservation equations $T^{\mu}_{\phantom{\mu}t;\mu} = 0$
and $T^{\mu}_{\phantom{\mu}\phi;\mu} = 0$ can be cast as
$\Omega_{,r}A_{\phi,\theta} = \Omega_{,\theta}A_{\phi,r}$ and
$(\sqrt{-g}F^{\theta r})_{,r}A_{\phi,\theta} = (\sqrt{-g}F^{\theta
r})_{,\theta}A_{\phi,r}$.
It is obvious that $\Omega$ and $\sqrt{-g} F^{\theta r}$ are
functions of $A_\phi$, i.e., $\Omega \equiv \Omega(A_\phi)$ and
$\sqrt{-g}F^{\theta r} \equiv I(A_\phi)$, where $\Omega$ and $I$ are
as-yet unspecified functions. Substitute Equations
(\ref{poloidal}), (\ref{electricfield}), (\ref{toroidal}), and the
relation $F^{\theta r} = I/\sqrt{-g}$ into the equation $F^{\theta
r} = g^{\theta\mu} g^{r\nu} F_{\mu\nu}$, we can readily arrive at
\begin{\eq}\label{znajek}
B^\phi = - \frac{I \Sigma + (2 \Omega r - a) \sin\theta
A_{\phi,\theta}} {\Delta \Sigma \sin^2\theta} \ ,
\end{\eq}
which relates the toroidal magnetic field $B^{\phi}$ to the
functions $A_{\phi}(r,\theta)$, $\Omega(A_{\phi})$, and
$I(A_{\phi})$.
To determine the unknown
functions of $\Omega(A_{\phi})$ and $I(A_{\phi})$, the remaining
momentum conservation equations in the $r$ and $\theta$ direction
$T^{\mu}_{\phantom{\mu}r;\mu} = 0$ and
$T^{\mu}_{\phantom{\mu}\theta;\mu} = 0$ have to be considered in
greater details. The two conservation equations in the $r$ and
$\theta$ directions are equivalent and read
\begin{\eq}\label{grad-shafranov}
-\Omega \left[(\sqrt{-g}F^{tr})_{,r} +
(\sqrt{-g}F^{t\theta})_{,\theta} \right] + F_{r\theta}I'(A_\phi) +
\left[(\sqrt{-g}F^{\phi r})_{,r} +
(\sqrt{-g}F^{\phi\theta})_{,\theta} \right] = 0 \ ,
\end{\eq}
where the prime denotes derivative with respect to $A_{\phi}$.
Note that above equation is equivalent to Equation (3.14) in (\cite{Blandford1977d}),
which is also widely called  Grad-Shafranov equation. The three
functions $A_{\phi}(r,\theta)$, $\Omega(A_{\phi})$, and
$I(A_{\phi})$ are related
by the nonlinear equation (\ref{grad-shafranov}).

\section{Collimated Jet Solutions -- A Perturbative Approach}
Since the Grad-Shafranov equation is highly nonlinear, our
strategy is to find its solution in the simplest case for
non-rotating black holes, i.e., $a=0$ and then to perturb the
simplest solution by allowing the black hole's spin, $a$, to
increase slowly. Namely, the corresponding solution can be
expressed,  up to $O(a^2)$, as
\begin{eqnarray}
A_{\phi}&=&A^{(0)}_{\phi}+a^2A^{(2)}_{\phi}+O(a^4) \ , \\
\Omega&=&a\Omega^{(1)}+O(a^3) \ , \\
B^{\phi}&=&a B^{\phi(1)}+O(a^3) \ .
\end{eqnarray}
Keep in mind that $\Omega$ and $\sqrt{-g}F^{\theta
r}$ are both functions of $A_{\phi}$, and they should be in the
form of
\begin{\eq}
\Omega = \Omega(A_\phi)  = a \omega(A_\phi) \ , \quad
\sqrt{-g}F^{\theta r}= I(A_\phi)=a i(A_{\phi}) \ .
\end{\eq}
We now consider the zeroth-order solution $A^{(0)}_{\phi}$ with
$a=0$. The simplest force-free field around non-rotating black
holes is actually the potential field in the Schwarzschild
spacetime (\cite{Ghosh2000}). In this case $\Omega(r,\theta) = 0$
and $B^\phi(r,\theta) = 0$. The non-vanishing components of the
Faraday tensor $F_{\mu\nu}$ are $F_{r\phi},F_{\phi
r},F_{\theta\phi},F_{\phi\theta}$. It is easy to know that
Equation (\ref{energymomentum}) holds automatically for the
$\mu=t,\phi$ components, and the $\mu=r,\theta$ component
equations give the identical result as follows:
\begin{\eq}
\mathcal{L} A^{(0)}_\phi \equiv \left\{\frac{1}{\sin \theta}
\frac{\partial}{\partial r} \left(1-\frac{2}{r}\right)
\frac{\partial}{\partial r} + \frac{1}{r^2}
\frac{\partial}{\partial \theta} \frac{1}{\sin
\theta}\frac{\partial}{\partial \theta} \right\} A^{(0)}_{\phi} =
0 \ .
\end{\eq}
There exists a zeroth-order collimated, uniform magnetic field
solution
\begin{\eq}
A^{(0)}_{\phi} = r^2\sin^2\theta \ .
\end{\eq}
The field line of this solution is of a highly collimated
cylindrical shape. If the black hole is spinning, toroidal
magnetic fields will be generated, and the magnetic cylinder will
be twisted and turns into a helically twisted structure. Note that
different zeroth-order solutions, i.e., the monopole solution and
the paraboloidal solution, were adopted in
(\cite{Blandford1977d}). The explicit dependence of zeroth-order
solutions on the coordinate $r$ and $\theta$ are displayed in
Table 1.

According to Equation (\ref{znajek}), we find
\begin{\eq}\label{bphi}
B^{\phi(1)} = \frac{2 \cos \theta (1-2r \omega(A_\phi)) -
i(A_\phi)/\sin^2\theta}{r^2-2r} \ .
\end{\eq}
If we require $B^{\phi}$ to be well-behaved on the horizon (Znajek
horizon condition \cite{Znajek1977b}), then $r=2$ must be a root to the equation, $2
\cos\theta (1-2r \omega) - i(A_\phi)/\sin^2\theta=0$, namely, $i(4
\sin^2\theta) = 2\cos\theta\sin^2\theta \left[ 1 -
4\omega(4\sin^2\theta) \right]$. This equation can be written in a more
compact form as $ i(x)= \sqrt{4-x}\left(\frac{1}{4}-\omega(x)\right) x
$, where $x = 4\sin^2\theta$. Since $i=i(A_{\phi})$ and this
equation can be written as
\begin{\eq} \label{h-Omega}
i(A_\phi) = \sqrt{4 - A_\phi} \left( \frac{1}{4}-\omega(A_\phi) \right)
A_\phi \ .
\end{\eq}

Before diving into solving the GS equation, it is helpful to
analyze the behavior of energy flux first. The energy flux is
defined as (\cite{Damour1975b}\cite{Znajek1977b})
\begin{\eq}
T^{r}_{\phantom{x}t} = F^{r\theta}F_{t\theta} =
-\frac{1}{\sqrt{-g}} I(A_\phi)\Omega(A_\phi) A_{\phi,\theta} \ ,  \
T^{\theta}_{\phantom{x}t} = F^{\theta r}F_{tr} =
\frac{1}{\sqrt{-g}} I(A_\phi)\Omega(A_\phi) A_{\phi,r} \ .
\end{\eq}
On the cylinder surface, $r \sin\theta = 2$, it is easy to know
that $T^{r}_{\phantom{x}t} \equiv 0$ and
$T^{\theta}_{\phantom{x}t} \equiv 0$. So $r \sin\theta = 2$ serves
as a boundary that no energy flux penetrates. In the outer region,
$r\sin\theta > 2$,
we can simply choose
\begin{\eq}
I(A_\phi) = 0 \  \ \Omega(A_{\phi}) = 0 \quad (r\sin\theta > 2),
\end{\eq}
which make sure that $T^r_{\phantom{x} t} = T^\theta_{\phantom{x}
t} \equiv 0$ in the outer region. In the following we will see
that our choice naturally makes the global solution continuous
across the interface at $r\sin\theta = 2$.

\noindent With some tedious manipulations, Equation
(\ref{grad-shafranov}) can be reduced, accurate to order $O(a^2)$,
to
\begin{\eq}\label{Aphi2nd}
  \mathcal{L} A^{(2)}_{\phi}= S(r,\theta) = \left\{\begin{array}{ll}
       S_{\mathrm{in}}(r,\theta) \  & \mbox{if}\quad r\sin\theta < 2 \\
       S_{\mathrm{out}}(r,\theta) \  & \mbox{if}\quad r\sin\theta > 2 \\
       \end{array} \right.  \ ,
\end{\eq}
where the source term in the inner region ($r \sin\theta < 2$) is
\[
S_{\rm in}(r, \theta) = 
8\sin\theta \cos^2\theta/r^3 
-6\sin\theta\cos^2\theta/r^2 
+ (1 - 2\omega r)(\sin^2\theta B^{\phi(1)})_{,\theta}
\]
\[
+ 4 \sin\theta \omega^2 r^2  
+ 4 \sin\theta(3\cos^2\theta-1) \omega^2 r 
+ r^2\sin\theta B^{\phi(1)} i'
\]
\begin{\eq}\label{Sinner}
- 8 \omega'r \sin^5\theta + 4 r^3\sin^3\theta \omega \omega'\left( r +
2\cos^2\theta \right) \ ,
\end{\eq}
and the source term in the outer region ($r \sin\theta > 2$) is
\begin{\eq}
S_{\rm out}(r,\theta) = \frac{8 \sin\theta \cos^2\theta}{r^3} -
\frac{2 \sin\theta}{r^2} + \frac{4 \sin\theta}{r^2(r-2)} (3
\cos^2\theta-1) \ .
\end{\eq}
Note that the prime in the above equations represents the
derivative with respect to $A_{\phi}$.
The Znajek horizon condition imposes the equation (\ref{h-Omega})
between $I$ and $\Omega$. To specify the collimated jet, we still
need to know the behavior of angular velocity of the magnetic
field, $\Omega(A_{\phi})$. Usually we need to solve the above
inhomogeneous Grad-Shafranov equation before we get the angular
velocity of the magnetic field. Fortunately, we find that the
convergence condition can be applied to get the further details
about this solution.
According to BZ77, the sufficient and necessary condition for the
existence of convergent solution of $A_\phi^{(2)}$ is that the
integral $\int_2^\infty dr\int_0^{\pi} d\theta
\frac{|S(r,\theta)|}{r}$ converges (convergence condition). It is
easy to prove that the contribution from the outer region
\begin{\eqn}
\int_2^\infty dr\int_\delta ^{\pi - \delta} d\theta \frac{|S(r,
\theta)|}{r} = \int_2^\infty dr\int_\delta ^{\pi - \delta} d\theta
\frac{|S_{\rm out}(r, \theta)|}{r}
\end{\eqn}
converges, where $\delta = \arcsin (2/r)$. So we only need to
require the contribution from inner region
\begin{\eq}
\int_2^\infty dr \int_0^\delta d\theta \frac{|S(r, \theta)|}{r} =
\int_2^\infty dr \int_0^\delta d\theta \frac{|S_{\rm
in}(r,\theta)|}{r} \
\end{\eq}
to be convergent. Assuming $\omega,\omega'\sim O(1)$, the following three
source terms in Equation (\ref{Sinner})
\begin{\eq}
4\omega^2r^2\sin\theta \sim 4 \omega \omega'r^4\sin^3\theta \sim  r^2 \sin\theta
B^{\phi(1)} i'
\end{\eq}
are of the same order $O(r)$ and will lead to logarithm divergence of the integral above, 
where we have used the fact that $0 \le \theta \le \delta \sim O(1/r)$ for the inner region. 
Note that except the above three terms, the
contribution from all other terms in Equation (\ref{Sinner}) is
convergent and are not listed here. So the convergence condition
requires
\begin{\eqn}
4 \omega^2 r^2 \sin\theta + 4 \omega \omega^{\prime} r^4 \sin^3\theta +
r^2\sin\theta B^{\phi(1)} i^{\prime} = 0  \ .
\end{\eqn}
Accurate to $O(r)$, the above equation can be written equivalently
as
\begin{\eqn}
4\omega^2 A_\phi + 4 \omega \omega' A_\phi^2 - \frac{(i^2)'}{2} = 2
(\omega^2A_\phi^2)' - \frac{(i^2)'}{2} = 0 \ ,
\end{\eqn}
where we have used the result of Eq.(\ref{bphi}).
Obviously the above equation can be integrated as
\begin{\eq}\label{constraint2nd}
{\rm const}=4\omega^2A_\phi^2 - i^2 = 4\omega^2A_\phi^2 - (4-A_\phi)\left(
\frac{1}{4}-\omega\right)^2 A_\phi^2 \ .
\end{\eq}
Note that $A_{\phi} =0$ at the polar axis, hence the integration
constant vanishes.
This equation constitutes a quadratic equation for the unknown
function $\omega=\omega(A_{\phi})$, which can be solved explicitly as,
\begin{\eq}
\omega = \frac{\sqrt{4 - A_\phi}}{4 \left(2 + \sqrt{4 - A_\phi}
\right)} \quad{(r\sin\theta < 2)} \ ,
\end{\eq}
which is consistent with the result of \cite{Beskin2009MHD} 
who obtained the same solution using a different approach.
Using Eq.(\ref{h-Omega}), we get 
\begin{\eq}
i = \frac{\sqrt{4 - A_\phi} A_\phi}{2 \left(2 + \sqrt{4 - A_\phi}.
\right)}\quad{(r\sin\theta < 2)} \ .
\end{\eq}
Surely we may explicitly write $A^{(2)}_\phi$ as sum of infinite
series (\cite{Blandford1977d}). But we do not plan to do that,
because we have obtained all quantities that are of physical
interests even not knowing the details about $A^{(2)}_\phi$,
which only provides information about the distortion of magnetic field lines in $r-\theta$ plane.

\section{Physical Properties of Collimated, Magnetically Dominated Jets}
We have obtained the explicit analytical expressions of $\omega =
\omega(A_{\phi})$ and $i = i(A_{\phi})$. Some interesting physical
properties of this collimated jet solution can be further
explored, such as the energy extraction rate, the energy extraction efficiency,
the stability of the jet solution, the comparison with solutions of BZ77 and numerical simulations.

The energy extraction rate is defined as $ \dot E = - 2
\pi \int_0^\pi\sqrt{-g} T^{r}_{\phantom{x}t}d\theta = 2\pi
\int_0^{\pi} I(A_\phi)\Omega(A_\phi) dA_\phi$. Direct integration leads to
\begin{\eqn}
\dot E/2
=  2\pi \int_0^4 I \Omega d A_\phi
= 8\pi a^2\left(2 - 2\ln2 - \frac{7}{12} \right) \approx 0.24 \pi
a^2 \ ,
\end{\eqn}
where the factor $2$ on the left hand side is to include
contributions from both hemispheres. The comparison between
previous analytic solutions and the jet solution is listed in
Table 1. 
In the table, the energy
extraction efficiency, $\bar \epsilon$, is defined as \citep{Blandford1977d}
\begin{\eq}
\bar\epsilon = \frac{\int I\Omega dA_\phi}{\int I\Omega_H dA_\phi} \ .
\end{\eq}
All the zeroth-order solutions are normalized to keep the amount
of magnetic flux crossing the horizon identical.  It is clear that
the collimated jet power is reduced by a factor about 30\%
compared to the split monopole jet model, which is more consistent
with recent simulation results \cite{Penna2013a}. 


\begin{table}
\caption{Comparison between our collimated jet solution and prior
solutions. We have normalized all solutions by requiring magnetic flux threading
BH event horizon to be unity. Where $f(r,\theta)=\frac{1}{4\ln 2}\left[r(1-\cos\theta)
+ 2(1+\cos\theta)(1-\ln(1+\cos\theta))-4(1-\ln2)\right]$, and $g(A_\phi)$ varies from
$0.5$ to $0.265$ with $f(r,\theta)$ varying from $0$ to $1$, see \citet{Blandford1977d}.}
\begin{tabular}{m{3cm} m{2cm} m{3cm} m{3cm} m{3cm} m{2cm}}
\hline \hline
{\rm solution}&$A_\phi^{(0)}$ & $\Omega$ & $I(A_\phi)$ &$\dot E$& $\bar\epsilon$\\
\hline
{\rm Split Monopole}& $-\cos\theta$& $\Omega_H/2$& $\Omega(1-A_\phi^2)$ &$0.67\pi \Omega_H^2$&  $0.50$\\
{\rm Paraboloidal}  & $f(r,\theta)$ & $\Omega_H g(A_\phi)$ & $2\Omega A_\phi$ & $ 0.55\pi \Omega_H^2$ & 0.38\\
{\rm Collimated Jet}& $r^2\sin^2\theta/4$& $\Omega_H\frac{\sqrt{1-A_\phi}}{1+\sqrt{1-A_\phi}}$& $2\Omega A_\phi$ &$0.48\pi \Omega_H^2$&  $0.36$\\
\hline
\end{tabular}
\end{table}

In Fig.1, we show the variation the angular velocity $\Omega$ and total electric current $I(A_\phi)$ on the horizon,
which matches numerical simulation results, 
such as the bottom panel of Fig.5 of \cite{Komissarov2005} .
\citet{Komissarov2005} noticed a sharp transition between the
rotating jet column of magnetic field lines penetrating the black
hole horizon and the non-rotating field lines which are not
attached to the black hole, and they interpreted the sharp
transition as a discontinuity smeared by numerical viscosity.
Our analytic solution clearly
shows that the transition is only a sharp turn instead
of a discontinuity, which is also confirmed by recent simulations
of higher resolution \cite{Alic2012}.

\begin{figure*}
\begin{center}
\includegraphics[scale=0.4]{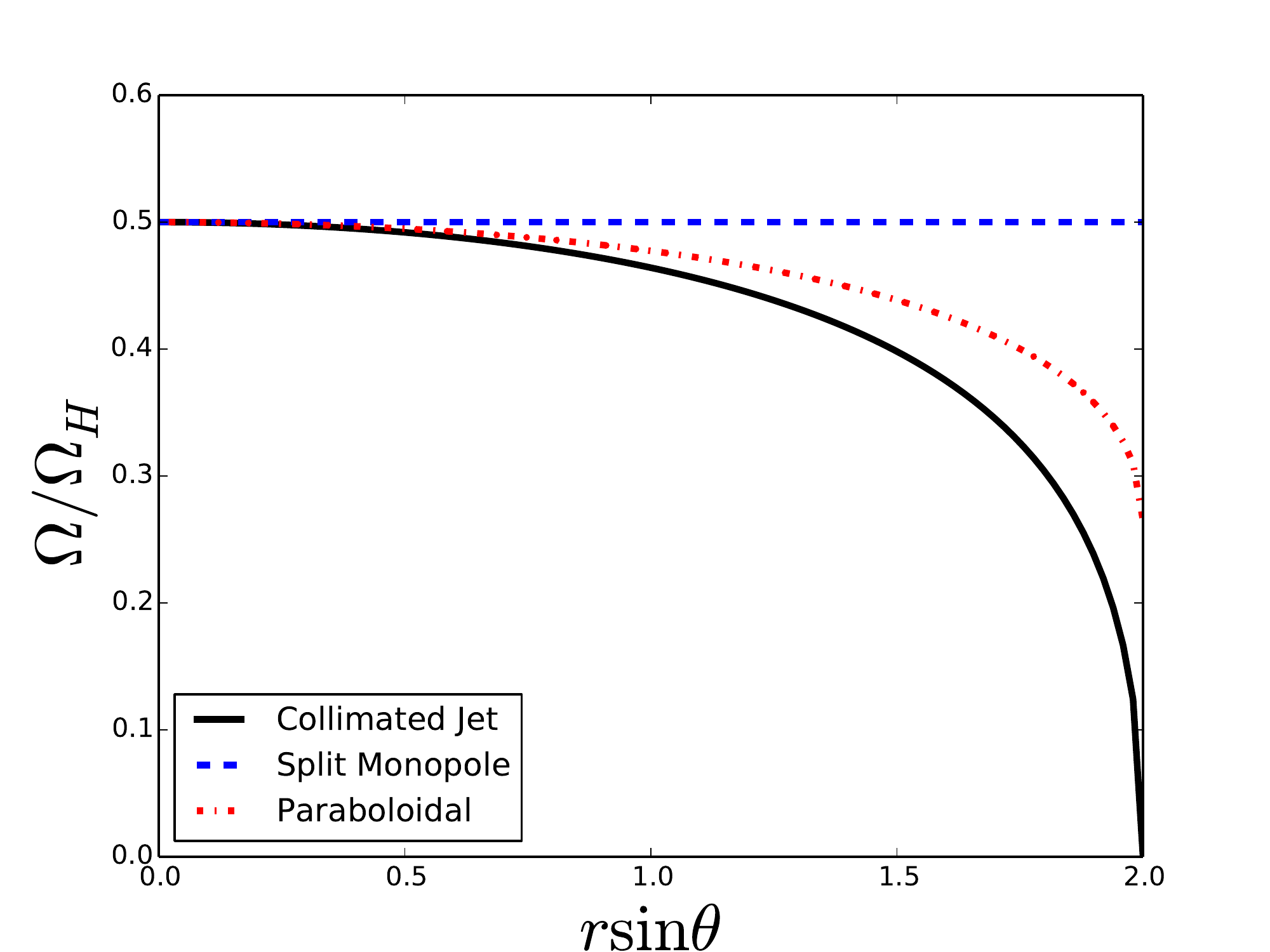}%
\includegraphics[scale=0.4]{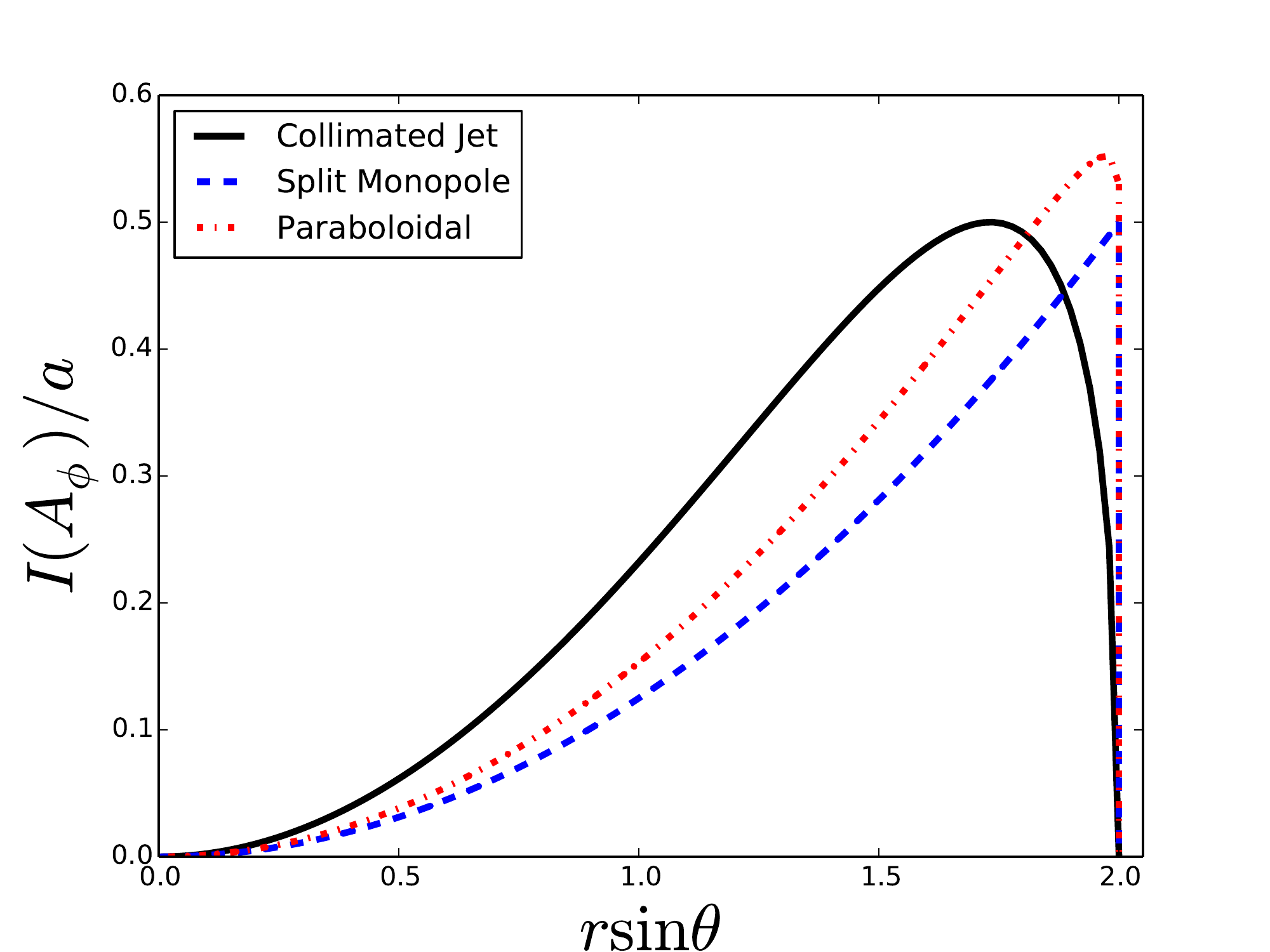}
\caption{ The comparison of three solutions: the variation
of the angular velocity $\Omega$ and the total electric current $I$ on horizon $r=2$. }
\end{center}
\end{figure*}

The stability of jets is also an issue of vital astrophysical
importance. Time dependent GRMD simulations of black hole
magnetospheres have demonstrated the stability of split monopole
solution (\cite{Komissarov2001}) and the collimated jet solution
(\cite{Komissarov2005}). Recent calculations show that mode growth
rates are far lower than those predicted by the Kruskal-Shafranov
stability criterion, suggesting that it may not be appropriate for
jet stability analysis (\cite{Narayan2009g}). Thus we adopt the
the criterion proposed by \cite{Tomimatsu2001}, since it
explicitly accounts for the effects of field line rotation. This
criterion states that the magnetosphere will possibly be unstable
only when $|\hat B_\phi/\hat B_z| > \Omega r\sin\theta \ ,$
where ${\hat B_\phi}$ and ${\hat B_z}$ are local toroidal and
poloidal field in the zero-angular-momentum observers' frame. To
be specific, accurate to $O(a)$, the toroidal and poloidal
magnetic field are
\begin{\eq}
|\hat B_\phi| \simeq
\frac{I}{r\sin\theta} = 2\Omega r\sin\theta \ , \quad |\hat B_z| =
2 \ ,
\end{\eq}
respectively. The ratio of the two is approximately,
$| \hat B_\phi/\hat B_z | =\Omega  r\sin\theta \ .$
It is clear that the instability threshold is not satisfied and
the collimated jet solution is a stable solution.

\section{Summary and Discussion}
We present a general method for perturbative solutions of  Blandford-Znajek mechanism.
Assuming stationary axisymmetric and force-free magnetospheres,
the energy-momentum equations can be divided into 2 constraint equations $\Omega \equiv \Omega(A_\phi)$ and
$\sqrt{-g}F^{\theta r} \equiv I(A_\phi)$  and GS equation (Eq.\ref{grad-shafranov}),
where GS equation is a second order differential equation which requires two boundary conditions.
We propose the horizon regularity condition and the convergence condition as the two boundary conditions.
With the boundary conditions, all physical quantities such as the angular velocity of magnetic fields $\Omega$,
the total current $I$ and energy extraction rate $\dot E$ are self-consistently spicified.

As an example, we construct a highly-collimated and magnetically-dominated jet
solution in the vicinity of spinning black holes . The nonlinear
GS equation (\ref{grad-shafranov}) is investigated analytically to
get axisymmetric steady-state force-free jet solutions. This
equation is solved by a perturbation technique. In this paper we
choose a uniform and collimated zeroth-order solution,
$A^{(0)}_{\phi} = r^2\sin^2\theta$, in the Schwarzchild spacetime.
The higher order solution, $A^{(2)}_{\phi}$, which accounts for
the effect of black hole spin, can be obtained based on the
zeroth-order solution around nonrotating black holes. It is
straightforward yet tedious to get the solution of inhomogeneous
GS equation (\ref{Aphi2nd}) for $A^{(2)}_\phi$. Nevertheless,
$A^{(2)}_{\phi}$ can not provide further details about the
physical properties of the jet, such as the angular velocity of
the field lines, $\Omega$, and the poloidal current, $I$, we take
an alternative way to get these important physical quantities.
According to the Znajek horizon boundary condition, we know that
the angular velocity of the magnetic field, $\Omega$, and poloidal
current, $I$, are closely related by Equation (\ref{h-Omega}). The
condition for the existence of convergent solution to the
inhomogeneous GS equation (\ref{Aphi2nd}) imposes another
constraint between $\Omega$ and $I$, i.e., Equation
(\ref{constraint2nd}). With these two equations, $\Omega$ and $I$
could be explicitly determined. It is clear that even we do not
get the explicit expression of the second-order solution
$A^{(2)}_{\phi}$, we still self-consistently determine all the
valuable physical variables.

Based on the known the angular velocity and toroidal magnetic
field of the collimated jet, we further explore the physical
properties of the jet solution, such as the jet power, energy
extraction rate. It is found that, given the magnetic field flux
threading the black hole horizon, the power of the highly
collimated jet is about 30\% lower than the standard split
monopole jet power and our solution is more consistent with recent
GRMHD simulation results. 

The jet field lines are helically twisted by the black hole's
spin. It has been shown the KS criterion for screw instability may
not be valid for rotating jet stability analysis and we adopt a
new criterion proposed by \cite{Tomimatsu2001}. We further study
how the effects of field line rotation influence the jet
stability. We find the rotation tends to stabilize the jet.

\begin{acknowledgments}
CY thanks the support by the National Natural Science
Foundation of China (grants 11173057 and 11373064), Yunnan Natural
Science Foundation (Grant 2012FB187). Part of the computation was
performed at HPC Center, Yunnan Observatories, CAS, China.
\end{acknowledgments}

\bibliography{ms}

\end{document}